\documentclass[aps,epsfig,twocolumn]{revtex4}
\usepackage{amssymb}
\usepackage{amsfonts}
\usepackage{graphicx}

\begin{document}

\title{Extending Hudson's theorem to mixed quantum states}
\author{A.~Mandilara$,^{1,2}$ E.~Karpov$,^{1}$ and N.~J.~Cerf$^{1,3}$}
\affiliation{$^{1}$Quantum Information and Communication, \'{E}cole Polytechnique,
Universit\'{e} Libre de Bruxelles, CP~165, 1050 Brussels, Belgium\\
$^{2}$Laboratoire Aim\'{e} Cotton, CNRS, Campus d'Orsay, 91405 Orsay, France%
\\
$^{3}$Research Laboratory of Electronics, Massachusetts Institute of
Technology, Cambridge, Massachusetts 02139}

\begin{abstract}
According to Hudson's theorem, any pure quantum state with a positive Wigner
function is necessarily a Gaussian state. Here, we make a step towards the
extension of this theorem to mixed quantum states by finding upper and lower
bounds on the degree of non-Gaussianity of states with positive Wigner
functions. The bounds are expressed in the form of parametric functions
relating the degree of non-Gaussianity of a state, its purity, and the
purity of the Gaussian state characterized by the same covariance matrix.
Although our bounds are not tight, they permit us to visualize the set of
states with positive Wigner functions.
\end{abstract}

\maketitle

The Wigner representation of quantum states \cite{Wigner}, which is realized
by joint quasi-probability distributions of canonically conjugate variables
in phase space, has a specific property which differentiates it from a true
probability distribution: it can attain negative values. Among pure states,
it was proven by Hudson \cite{Hudson} (and later generalized to
multi-mode quantum systems by Soto et Claverie \cite{Soto}) that the only
states which have non-negative Wigner functions are Gaussian states \cite%
{foot}. The question that naturally arises \cite{Hudson} is whether this
theorem can be extended to mixed states, among which not only Gaussian
states may possess a positive Wigner function. A logical extension of the
theorem would be a complete characterization of the convex set of states
with positive Wigner function. Although this question can be approached by
using the notion of Wigner spectrum \cite{Werner}, a simple and operational
extension of Hudson's theorem has not yet been achieved due to the
mathematical complications which emerge when dealing with states with
positive Wigner functions \cite{Werner}.

Motivated by the increasing interest for non-Gaussian states in
continuous-variable quantum information theory (see, e.g., \cite{bookcerf})
and the need for a better understanding of the de-Gaussification procedures
for mixed states (see, e.g., \cite{de-Gau}), we attempt here an exploration
of the set of states with positive Wigner functions using Gaussian states as
a reference. More precisely, we consider the subset of such states that have
the same covariance matrix as a reference Gaussian state. We obtain a
partial solution to the problem, by analytically deriving necessary
conditions (bounds) on a measure of non-Gaussianity for a state to have a
positive Wigner function. This set of conditions bounds a region in a
three-dimensional space with coordinates being the purity of the state, the
purity of the corresponding Gaussian state, and the non-Gaussianity. As
intuitively expected, the maximum degree of non-Gaussianity increases with a
decrease in the purity of both the state and its Gaussian corresponding
state.\smallskip

Before deriving the main results of this paper, let us recall a convenient
representation of the trace of the product of two one-mode quantum states, $%
\rho $ and $\rho ^{\prime }$, in terms of the Wigner representation \cite%
{Schleich}, 
\begin{equation}
\mathrm{Tr}\left( \rho \rho ^{\prime }\right) =2\pi \int \int dx\, dp\,
W_{\rho }\left( x,p\right) W_{\rho ^{\prime }}\left( x,p\right),  \label{over}
\end{equation}%
where $W_{\rho }$ is the \emph{Wigner function} of the state $\rho $. For
example, the \textit{purity} of a state, $\mathbf{\mu }\mathcal{[}\rho ]=%
\mathrm{Tr}\left( \rho ^{2}\right) $, may be calculated with the help of
this formula. For a state with a \emph{Gaussian} Wigner function determined
by the covariance matrix $\gamma $ and displacement vector $\mathbf{d}$, the
purity is simply $\mu \left[ \rho _{G}\right] =\left( \det \gamma \right)
^{-1/2}$. The matrix elements of the covariance matrix of state $\rho $ are
defined as 
\begin{equation}
\gamma _{ij}=\mathrm{Tr}(\{(\hat{r}_{i}-d_{i}),(\hat{r}_{j}-d_{j})\}\rho )
\end{equation}%
where $\hat{\mathbf{r}}$ is the vector of quadrature operators $\hat{\mathbf{%
r}}=(\hat{x},\hat{p})^{T}$, $\mathbf{d}=\mathrm{Tr}(\hat{\mathbf{r}}\rho )$,
and $\{\cdot ,\cdot \}$ is the anticommutator. Note that we can put the
displacement vector to zero with no loss of generality since the purity (and
all quantities we will be interested in) does not depend on $\mathbf{d}$. We
will thus consider states centered on the origin in this paper.

Our aim is to derive bounds on the non-Gaussianity, i.e., on the
\textquotedblleft distance\textquotedblright\ between a state $\rho $ of
purity $\mathbf{\mu }[\rho ]$ possessing a positive Wigner function and the
Gaussian state $\rho _{G}$ determined by the same covariance matrix. While
there are different measures in the literature for quantifying the distance
between two mixed states, we have chosen to use a recently proposed one 
\cite{nG},
\begin{equation}
\delta \left[ \rho ,\rho _{G}\right] =\frac{\mathbf{\mu }[\rho ]+\mathbf{\mu 
}\left[ \rho _{G}\right] -2\, \mathrm{Tr}\left( \rho \rho _{G}\right) }{2%
\mathbf{\mu }[\rho ]}.  \label{delta}
\end{equation}
Although the quantity $\delta \lbrack \rho ,\rho _{G}]$ is obviously not
symmetric under the permutation of the two states, it is convenient for
quantifying the non-Gaussian character of $\rho $ in the sense that $\delta
\in \lbrack 0,\varepsilon ]$, with $\varepsilon <1$, and $\delta =0$ is
attained if and only if $\rho \equiv \rho _{G}$. For one-mode states, it is
conjectured in Ref.~\cite{nG} that $\varepsilon =1/2$.
\smallskip

In a first step, we are going to derive bounds on the trace overlap $\ 
\mathrm{Tr}\left( \rho \rho _{G}\right) $ for fixed values of $\mathbf{\mu }%
[\rho _{G}]$ and $\mathbf{\mu }[\rho ]$. It will then be straightforward to
express bounds on the non-Gaussianity $\delta \left[ \rho ,\rho _{G}\right] $
in terms of $\mathbf{\mu }[\rho _{G}]$ and $\mathbf{\mu }[\rho ]$ by using
Eq. (\ref{delta}).

We use Eq.(\ref{over}) in order to reformulate the problem as an
optimization problem that can be tackled with the method of Lagrange
multipliers. More specifically, we need to extremize the functional $%
I[W_{\rho }]=\mathrm{Tr}\left( \rho \rho _{G}\right) $ represented by Eq.~(%
\ref{over}) with the constraint that the Gaussian Wigner function $W_{\rho
_{G}}$ and the positive function $W_{\rho }$ possess the same second
moments. In order to simplify our derivation, we apply a symplectic
transformation $S$ on the states $\rho $ and $\rho _{G}$, giving
$S\rho S^{\dagger }$ and $S\rho _{G}S^{\dagger }$ respectively, in such a
way that the Gaussian state becomes invariant under rotation in the $x$-$p$
plane (i.e., becomes a thermal state). In this way the problem is reduced to
a simpler but equivalent one, since the functional $I[W_{\rho }]$ and the
purities of the states remain invariant under $S$ and since the positivity
of $W_{\rho }$ is preserved. This last statement can be justified by the
fact that the time evolution of a Wigner function under a quadratic
Hamiltonian can always be viewed as an affine transformation on the
variables $x$ and $p$ \cite{Schleich}. Furthermore, we claim that the
function $W_{\rho }^{ex}$ which extremizes the functional $I[W_{\rho }]$ is
invariant as well under rotation in the $x$-$p$ plane, and we will justify
this assumption at the end of the derivation.

After the application of the symplectic transformation and under the assumption
of rotation-invariant solutions, the functions $W_{\rho }(r)$ and
$W_{\rho_{G}}\left( r\right) =\frac{1}{2\pi C}\mathrm{e}^{-r/2C}$
only depend on the squared radius $r=x^{2}+p^{2}$,
and the functional $I[W_{\rho }]$ is written in a simpler form as 
\begin{equation}
I[W_{\rho }]=\mathrm{Tr}\left( \rho \rho _{G}\right) =2\pi
^{2}\int_{0}^{\infty }W_{\rho }\left( r\right) W_{\rho _{G}}\left( r\right)
dr.  \label{func2}
\end{equation}%
The constrains that we impose on the function $W_{\rho }$ can be summarized
as follows:

\begin{itemize}
\item[(1)] It is positive for the values of $r$ belonging to some set $\mathfrak{s%
}$ and zero elsewhere.

\item[(2)] It is normalized, $\pi \int_{\mathfrak{s}}W_{\rho }(r)dr=1$.

\item[(3)] It has the same variance as the corresponding Gaussian state, $\rho
_{G}$ 
\begin{equation}  \label{variance}
\pi \int_{\mathfrak{s}}W_{\rho }(r)rdr=2C=1/\mu \left[ \rho _{G}\right] .
\label{c3}
\end{equation}

\item[(4)] It is such that the state $\rho $ has purity $\mathbf{\mu }[\rho ]$, 
\begin{equation}  \label{pur}
2\pi ^{2}\int_{\mathfrak{s}}W_{\rho }^{2}(r)dr=\mathbf{\mu }[\rho ].
\label{c4}
\end{equation}

\item[(5)] It is square integrable and continuous.
\end{itemize}

This last requirement follows directly from the general property of Wigner
functions,
\begin{eqnarray}
\int_{-\infty }^{\infty }W(x,p)dp &=&\left\langle x\right\vert \rho
\left\vert x\right\rangle  \label{c7} %\\
%\int_{-\infty }^{\infty }W(x,p)dx &=&\left\langle p\right\vert \rho
%\left\vert p\right\rangle .  \label{c8}
\end{eqnarray}%
Recall that a state can always be diagonalized in a basis of pure states,
namely $\rho =\sum_{i}\lambda _{i}\left\vert \psi _{i}\right\rangle
\left\langle \psi _{i}\right\vert $. Since wave functions must satisfy the
conditions of continuity and integrability in both position and momentum
representation, one concludes that a Wigner function of variables $x$ and $p$%
, and more generally of any variable that is a continuous function on these,
e.g., $r=x^{2}+p^{2}$, has to satisfy the same requirements.

Finally, let us stress that without the requirement of positive definiteness
of the operator $\rho $, the set of conditions listed above is \textit{not
sufficient} to constrain the solutions $W_{\rho }(r)$ to eligible Wigner
functions. To our knowledge, there exists no operational criterion on
phase-space functions ensuring that the operator $\rho $ is physical (see 
\cite{Tata} for an extensive discussion). On the other hand, one can verify
whether a quasi-probability distribution is unphysical by using a theorem
which states that a square integrable and normalized function is an eligible
Wigner function if its overlap with the Wigner function of
\textsl{every} pure state is positive \cite{Hillery}.

After having applied the method of Lagrange multipliers, we obtain the
extremal solution 
\begin{equation}
W_{\rho }^{ex}(r)=A_{1}+A_{2}\frac{1}{2\pi C}\mathrm{e}^{-r/2C}+A_{3}r\,,
\label{form}
\end{equation}%
with the $A$'s being determined by conditions \textsl{2-4}. Square
integrability, condition \textsl{5}, limits the class of possible functions $%
W_{\rho }^{ex}(r)$ in Eq.~(\ref{form}) to those \ that have zero, one, or
two positive roots denoted as $r_{B}$ (in the one- and two-root cases) and 
$r_{A} $ (in the two-root case). Furthermore the condition of continuity
dictates that $\mathfrak{s}=\left[ r_{A},r_{B}\right] $ in the two-root case, $%
\mathfrak{s}=\left[ 0,r_{B}\right] $ in the one-root case, and
$\mathfrak{s}=\left[ 0,\infty \right] $ in the zero-root case.
The latter case is the trivial
one, where $W_{\rho }^{ex}(r)$ coincides with $W_{\rho _{G}}(r)$ and thus $%
\delta \left[ \rho ,\rho _{G}\right] $ vanishes. We treat the other two
cases separately, and obtain two continuously connected branches of
solutions for $W_{\rho }^{ex}$. The expressions that we obtain for $\mathrm{%
Tr}\left( \rho \rho _{G}\right) ^{ex}$ and $\mathbf{\mu }[\rho ]^{ex}$ are
highly non-linear, so that it is not possible to derive an analytic
expression that directly connects the two quantities. Nevertheless, we are
able to express the extremal solutions in the form of parametric functions.

(I) Two roots : $W_{\rho }^{ex}(r_{A})=W_{\rho }^{ex}(r_{B})=0.$
We express the extremum purity $\mathbf{\mu }[\rho ]^{ex}$ and overlap 
$\mathrm{Tr}\left( \rho \rho _{G}\right) ^{ex}$ in terms of the purity of the
corresponding Gaussian state $\mu \lbrack \rho _{G}]$ and parameter $\alpha
=(r_{B}-r_{A})\mu \lbrack \rho _{G}]$,

%TCIMACRO{\TeXButton{TeX field}{\begin{widetext}}}%
%BeginExpansion
%\begin{widetext}%
%EndExpansion
\begin{equation}
\mathbf{\mu }[\rho ]^{ex} =\mathbf{\mu }[\rho _{G}]\frac{2\left( \alpha
^{2}-9\sinh (\alpha )\alpha +2\left( \alpha ^{2}+6\right) \cosh (\alpha
)-12\right) }{3\alpha \left( \alpha \cosh \left( \frac{\alpha }{2}\right)
-2\sinh \left( \frac{\alpha }{2}\right) \right) ^{2}},  \label{IImain}
\end{equation}
\begin{eqnarray}
\mathrm{Tr}\left( \rho \rho _{G}\right) ^{ex} =&\mathbf{\mu }[\rho _{G}]2exp%
\left[ -\frac{\alpha \left( \alpha +e^{\alpha }(2\alpha -3)+3\right) }{%
3\left( e^{\alpha }(\alpha -2)+\alpha +2\right) }\right] \nonumber\\
 &\times\left( e^{\alpha
}-1\right) /\alpha, \label{IImainb} 
\end{eqnarray}

where $0<\mu \lbrack \rho _{G}]\leq 1$. By imposing the condition $r_{A}>0$,
we obtain the bound $0<\alpha \leq x_{r}$, with $x_{r}$ being the root of
equation,
\begin{equation}
e^{x}(x-3)+2x+3=0.  \label{lim}
\end{equation}

\begin{widetext}
(II) One root : $W_{\rho }^{ex}(r_{B})=0$. The extremal solution is
defined by the following pair of parametric functions, 
%TCIMACRO{\TeXButton{TeX field}{\begin{widetext}}}%
%BeginExpansion
%
%EndExpansion

%\bigskip 
\begin{eqnarray}
\mathbf{\mu }[\rho ]^{ex} &=&\mathbf{\mu }[\rho _{G}]\frac{4\left( e^{2\beta
}(\beta -3)^{2}+8e^{\beta }\beta (\beta -3)+\beta (\beta (2\beta
+9)+12)-9\right) }{\left( 2e^{\beta }(\beta -3)+\beta (\beta +4)+6\right)
^{2}},  \label{Imain} \\
\mathrm{Tr}\left( \rho \rho _{G}\right) ^{ex} &=&\mathbf{\mu }[\rho _{G}]%
\frac{4(\beta (\cosh (\beta )+2)-3\sinh (\beta ))}{2e^{\beta }(\beta
-3)+\beta (\beta +4)+6},  \label{Imainb}
\end{eqnarray}%
%TCIMACRO{\TeXButton{TeX field}{\end{widetext}} }%
%BeginExpansion
\end{widetext}
%EndExpansion
where $0<\mu \lbrack \rho _{G}]\leq 1$ \ and $\beta =r_{B}\mu \lbrack \rho
_{G}]$. The range of the latter parameter is $\beta \geq x_{r}$.

\begin{figure}[!b]
{\centering{\includegraphics*[ width=0.35\textwidth]{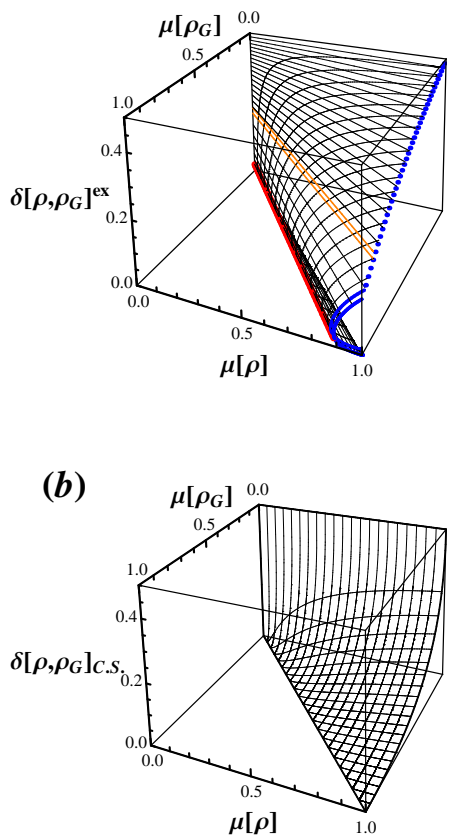}}}
%\vspace{0.1cm}
\caption{(Color Online) \textit{(a)} Upper bound on non-Gaussianity $\protect%
\delta \left[ \protect\rho ,\protect\rho _{G}\right] ^{ex}$ derived with the
Lagrange multipliers method. The double thin orange line marks
the boundary between the two branches of solutions.
The dotted blue line indicates the intersection with the plane
$\protect\mu \left[ \protect\rho \right] =1$
(also noted $\protect\delta ^{u.ult}$ in Fig. ~2) while the 
double thick blue line shows the intersection with the plane
$\protect\mu \left[ \protect\rho _{G}\right] =1$.
The red thick straight line denotes the `left' extremity of the surface. 
\textit{(b)} Lower bound
on non-Gaussianity $\protect\delta 
\left[ \protect\rho ,\protect\rho _{G}\right] _{CS}$
implied by the Cauchy-Schwarz inequality. We plot this
lower bound only up to the intersecting line of $\protect\delta \left[ 
\protect\rho ,\protect\rho _{G}\right] _{CS}$ and $\protect\delta \left[ 
\protect\rho ,\protect\rho _{G}\right] ^{ex}$ ($\protect\mu \left[ \protect%
\rho _{G}\right] =\protect\mu \left[ \protect\rho \right] $ and $\protect%
\delta \left[ \protect\rho ,\protect\rho _{G}\right] =0$). }
\label{Ubound}
\end{figure}

We now need to show that, although we have only considered solutions $%
W_{\rho }^{ex}$ with no angular dependence, our result is general. If we
waive this assumption and consider the most general case, we arrive to 
\begin{eqnarray}
W_{\rho }^{ex}(x,p) &=&A_{1}+A_{2}\frac{1}{2\pi C}\mathrm{e}^{-\left(
x^{2}+p^{2}\right) /2C}  \nonumber \\
&&+A_{3}x^{2}+A_{4}p^{2}+A_{5}xp,  \label{assym}
\end{eqnarray}%
which is the analog of Eq.~(\ref{form}) but allowing for an angular
dependence. We can then apply a phase rotation on states $\rho $ and $\rho
_{G}$ in order to eliminate the term $A_{5}xp$ in Eq.~(\ref{assym}). Such a
rotation does not affect the corresponding Gaussian state (since it is
thermal) nor the trace overlap, so that the resulting extremal function
becomes symmetric by reflection with respect to the $x$ or $p$ axis in phase
space. The conditions $\left\langle x^{2}\right\rangle =\left\langle
p^{2}\right\rangle =C$ \ for the function in Eq.~(\ref{assym}) is
satisfied if $A_{3}=A_{4}$. Thus, the most general solution
reduces to the rotation-invariant one, namely, Eq. (\ref{form}).

\begin{figure}[h]
{\centering{\includegraphics*[ width=0.3\textwidth]{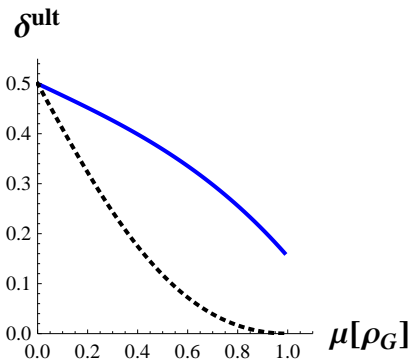}}} 
\vspace{0.1cm}
\caption{(Color Online) Ultimate  upper bound $\protect\delta^{u.ult}$ 
(blue solid line) on the non-Gaussianity of states with positive
Wigner fuction as a function of the putity of corresponding Gaussian state
$\mu[\rho_G]$. A lower estimate $\protect\delta^{l.ult}$
(black dashed line) on the ultimate upper bound 
that was obtained for a mixture of coherent states. The two curves limit the
region where the tight ultimate upper bound on non-Gaussianity must be
located.}
\label{Dbound}
\end{figure}

By using the derived bounds on the trace overlap 
[Eqs.~(\ref{IImain})--(\ref{Imainb})],
we plot in Fig.~\ref{Ubound}(a) the corresponding (upper) bounds
on the non-Gaussianity $\delta \left[ \rho ,\rho _{G}\right] ^{ex}$. By
direct inspection, we conclude that the intersection of the plotted surface
with the plane of pure states $\mathbf{\mu }[\rho ]=1$ provides us with an
upper bound on the non-Gaussianity of any state with a positive Wigner
function and fixed covariance matrix (or, equivalently, fixed
$\mathbf{\mu }[\rho _{G}]$) which is independent of its purity $\mathbf{\mu }[\rho ]$.
We denote it as the \textit{ultimate upper bound}
$\delta ^{u.ult}\left( \mathbf{\mu}\left[ \rho _{G}\right] \right) $
and its parametric expression can be
directly derived by setting $\mathbf{\mu }^{ex}[\rho ]=1$ in 
Eqs.~(\ref{IImain})--(\ref{Imainb}). In Fig.~\ref{Dbound} we plot
$\delta ^{u.ult}$ together
with a lower estimation on this, $\delta ^{l.ult}$, obtained by the convex
combination of two symmetrically displaced coherent states. The tight
ultimate upper bound on $\delta \left[ \rho ,\rho _{G}\right] $ must be
located between these two curves. 

The bounds derived by using the Lagrange multipliers method confine $\delta $
only from above. In order to obtain a lower bound on $\delta $, we need to
find an upper bound on the trace overlap. We achieve this by applying the
Cauchy-Schwarz inequality on the Wigner representation of the trace overlap,
[Eq.~(\ref{over})]. By using the definition of the purity, we arrive at 
\begin{equation}
\mathrm{Tr}\left( \rho \rho _{G}\right) \leq \sqrt{\mathbf{\mu }\left[ \rho
_{G}\right] \mathbf{\mu }\left[ \rho \right] }\equiv \mathrm{Tr}\left( \rho
\rho _{G}\right) _{C.S.}  \label{CS}
\end{equation}%
where $CS$ stands for \textquotedblleft Cauchy-Schwarz,\textquotedblright .
This bound, displayed in Fig.~\ref{Ubound}(b), delimits together with the
upper bound of Fig.~\ref{Ubound}(a), the region accessible for states with positive
Wigner function in this 3D representation. Let us note that
this lower bound holds for states with both positive
and negative parts Wigner functions.
\smallskip

Let us now address the question of the physicality of the upper bound,
namely the extremal solution $W_{\rho }^{ex}$. By resorting to Hudson's
theorem, we can conclude that the intersections of the surface with the
planes $\mathbf{\mu }[\rho ]=1$ and $\mathbf{\mu }[\rho _{G}]=1$ 
[see single and double blue lines in Fig.~\ref{Ubound}(a))]
cannot correspond to physical states where $\delta \neq 0$.
The only physical solution belonging to these
lines is thus one point, namely,
$\mathbf{\mu }[\rho ]=1$, $\mathbf{\mu }[\rho _{G}]=1$, and $\delta=0 $.
In order to test the physicality of the rest of the surface, we
applied the theorem mentioned above employing the eigenstates of the quantum
harmonic oscillator as test pure states. From our analytical results on the
first 40 number states, we infer that the only functions $W^{ex}_\rho$ 
giving a positive overlap with every number state as $n\rightarrow \infty $
are the states with $\mathbf{\mu }[\rho _{G}]=0$, that is,
infinitely mixed states.
Therefore, we conclude that the extremal solution of the form of 
Eq.~(\ref{form}) is unfortunately unphysical;
hence, \textit{our bound is not tight}.

Finally, one may notice in Fig.~\ref{Ubound}(a) that the left extremity of
the bound (red thick straight line) is on the left of the plane $\mu \lbrack \rho _{G}]=\mu
\lbrack \rho ]$. The equation for this line can be easily derived, 
\begin{equation}
\mu \lbrack \rho ]=\frac{8}{9}\mu \lbrack \rho _{G}]  \label{boundP}
\end{equation}%
and thus sets a lower bound on the purity of a mixed state given the purity
of the corresponding Gaussian state. This bound has been derived in
another context by Bastiaans \cite{Bast83} and has been proven to be the
asymptotic form of an exact expression derived later by Dodonov and Man'ko  
\cite{Man'ko} in the context of purity bounded uncertainty relation.
The exact bound is more strict than the bound in Eq.~(\ref{boundP}),
and it is realized by positive Wigner functions \cite{Dodonov}.
This fact confirms
again that our bound is unphysical but it also gives some evidence about the
underlying link between Hudson's theorem and the Heisenberg uncertainty
principle.

In conclusion, we have found both upper and lower bounds on the
non-Gaussianity of mixed states with positive Wigner function. These bounds
only depend on the purity and covariance matrix of these states, and an
ultimate upper bound can be derived that does not even depend on the purity,
making it experimentally accessible. An open question remains to derive
tighter bounds for the non-Gaussianity. All our results apply to one single
mode, so another natural question would be to investigate the case of
several modes.

\smallskip

The authors thank Julien Niset for fruitful discussions. A.M. gratefully
acknowledges financial support from the Belgian National Fund for Scientific
Research. This work was carried out with the financial support of the
European Commision via projects COMPAS and QAP, the support of the Belgian
Rederal program PAI via the Photonics project, and the support of the
Brussels-Capital Region via projects CRYPTASC and Prospective Research for
Brussels.

%\bigskip 

\end{document}